# Charge trap layer enabled positive tunable $V_{fb}$ in β-Ga$_2$O$_3$ gate stacks for enhancement mode transistors

Dipankar Biswas,[1] Chandan Joishi,[1,2] Jayeeta Biswas,[1] Prabhans Tiwari,[1] and Saurabh Lodha[1,a]

1) Department of Electrical Engineering, Indian Institute of Technology Bombay, Mumbai, Maharashtra 400076, India

2) Department of Electrical and Computer Engineering, The Ohio State University, Columbus, OH 43210, U.S.A.

E-mail: slodha@ee.iitb.ac.in

(Dated: 11 May 2020)

## Abstract

*β*-Ga$_2$O$_3$ based enhancement mode transistor designs are critical for the realization of low loss, high efficiency next generation power devices with rudimentary driving circuits. A novel approach towards attaining a high positive flat band voltage ($V_{fb}$) of 10.6 V in *β*-Ga$_2$O$_3$ metal-oxide-semiconductor capacitors (MOSCAPs), with the ability to fine tune it between 3.5 V to 10.6 V, using a polycrystalline AlN charge trap layer has been demonstrated. This can enable enhancement mode operation over a wide doping range. Excellent $V_{fb}$ retention of ~97% for 10$^4$ s at 55 °C was exhibited by the gate stacks after charge trapping, hence reducing the requirement of frequent charge injection cycles. In addition, low gate leakage current density ($J_g$) for high negative gate voltages ($V_g$~-60 V) indicates the potential of this gate stack to enable superior breakdown characteristics in enhancement mode transistors.

[a]To whom correspondence should be addressed

A high theoretical breakdown field ($E_{br}$) of 8 MVcm$^{-1}$ resulting from its ultrawide band gap ($E_g$=4.8 eV), combined with a reasonable electron mobility ($\mu$=300 cm$^2$V$^{-1}$s$^{-1}$), enable high figures of merit in power switching for $\beta$-Ga$_2$O$_3$.[1,2] In addition, cost effective growth of bulk crystals with low defect density over a wide doping range by scalable melt based methods (Czochralski, float-zone, edge-defined film-fed growth, etc.) has led to significant interest in this material for next generation power electronics.[3–5]

Over the years, $\beta$-Ga$_2$O$_3$ based vertical devices such as Schottky diodes, trench metaloxide-semiconductor field-effect transistors and current aperture vertical electron transistors (CAVETs), as well as lateral devices in the form of metal-semiconductor field-effect transistors (MESFETs), metal-oxide-semiconductor field-effect transistors (MOSFETs) and modulation-doped field-effect transistors (MODFETs) have been demonstrated with promising performance.[6–12] For lateral FETs, although existing reports show good progress in the development of depletion mode transistors with acceptable ON-state current densities and high breakdown voltages ($V_{br}$), the lack of p-type doping has limited the advancement of enhancement mode (normally-off) transistors. These are critical for low power and high efficiency next generation high power switches.[11–13] Recently, device design technologies such as recessed-gate, wrap-gate fin arrays, unintentionally doped channels and ferroelectric gate dielectrics have emerged to realize normally-off transistors on $\beta$-Ga$_2$O$_3$.[14–17] To further increase the current density and make it comparable to existing GaN-based technologies, modulation of higher charge density channels using different gate dielectrics are being explored.[18] Owing to its high dielectric constant (~8), reasonable conduction band offset ($\Delta E_c$) of 1.5-1.7 eV and a high quality interface with $\beta$-Ga$_2$O$_3$, Al$_2$O$_3$ has been the current dielectric of choice for $\beta$-Ga$_2$O$_3$ MOSFETs.[18–20] Developing further, we have recently demonstrated an Al$_2$O$_3$/SiO$_2$ bi-layer stack to leverage the benefits of both Al$_2$O$_3$ (high dielectric constant) and SiO$_2$ (higher $\Delta E_c$ of ~3.6 eV) which enabled a high, positive flat band voltage ($V_{fb}$) of 3.25 V (in comparison to 0.74 V on using only Al$_2$O$_3$).[21] The higher $V_{fb}$ was attributed to the negative oxide charge in SiO$_2$ deposited by atomic layer deposition (ALD).[22] The idea of utilizing negative gate dielectric charge to realize normally-off transistors is worth exploring in greater detail.

In this letter we realize normally-off operation through the introduction of a charge trap layer to house the negative charge in the gate dielectric stack. We demonstrate high positive $V_{fb}$ that can be tuned from 3.5 V to 10.6 V (with negligible impact on the dielectric/channel interface) in β-Ga$_2$O$_3$ MOSCAPs by controlled electron injection into a polycrystalline AlN charge trap layer sandwiched between Al$_2$O$_3$ and SiO$_2$. Moreover, excellent $V_{fb}$ retention of ~97% for 10$^4$ s at 55 °C helps reduce the frequency of charge injection cycles needed to sustain the charge. The introduction of AlN gives an added benefit of reducing the $J_g$ for high negative $V_g$ (~-60 V) compared to an Al$_2$O$_3$/SiO$_2$ bilayer stack. Hence, this stack design is expected to aid realization of normally-off transistors with excellent breakdown characteristics.

To study the morphology of AlN on β-Ga$_2$O$_3$ ($\bar{2}$01), 18 nm of the film was deposited by plasma-enhanced ALD (PEALD) using trimethylaluminum (TMA) and ammonia plasma at 200 °C. Grazing-incidence x-ray diffraction (GIXRD) was performed using Cu K$\alpha$ radiation in a Rigaku Smartlab diffractometer. The existence of diffraction peaks corresponding to (100), (101), (102), (110), (103) and (112) planes confirms the presence of hexagonal wurtzite phase in polycrystalline AlN as shown in Fig. 1(a).[23] X-ray photoelectron spectroscopy (XPS) analysis of the deposited AlN film (supplementary material) shows the presence of two different peaks in the Al2p spectrum at binding energies of 74.4 eV and 75.6 eV corresponding to Al-N and Al-O bonds, respectively. The presence of AlON is confirmed by the appearance of a sub-peak corresponding to Al-O-N bonds at 398.7 eV along with the main peak at 397.3 eV for Al-N bonds in the N1s spectrum.[24] Also, the atomic percentages extracted from the areas under the Al2p, N1s and O1s spectra show a distribution of 39.3%, 35.9% and 24.7% for Al, N and O, respectively. Despite having a similar Δ$E_c$ as that of Al$_2$O$_3$ with β-Ga$_2$O$_3$ (Fig. 1(b)), the deep energy level states present at the polycrystalline grain boundaries of AlN can serve as effective charge trapping sites.[24,25]

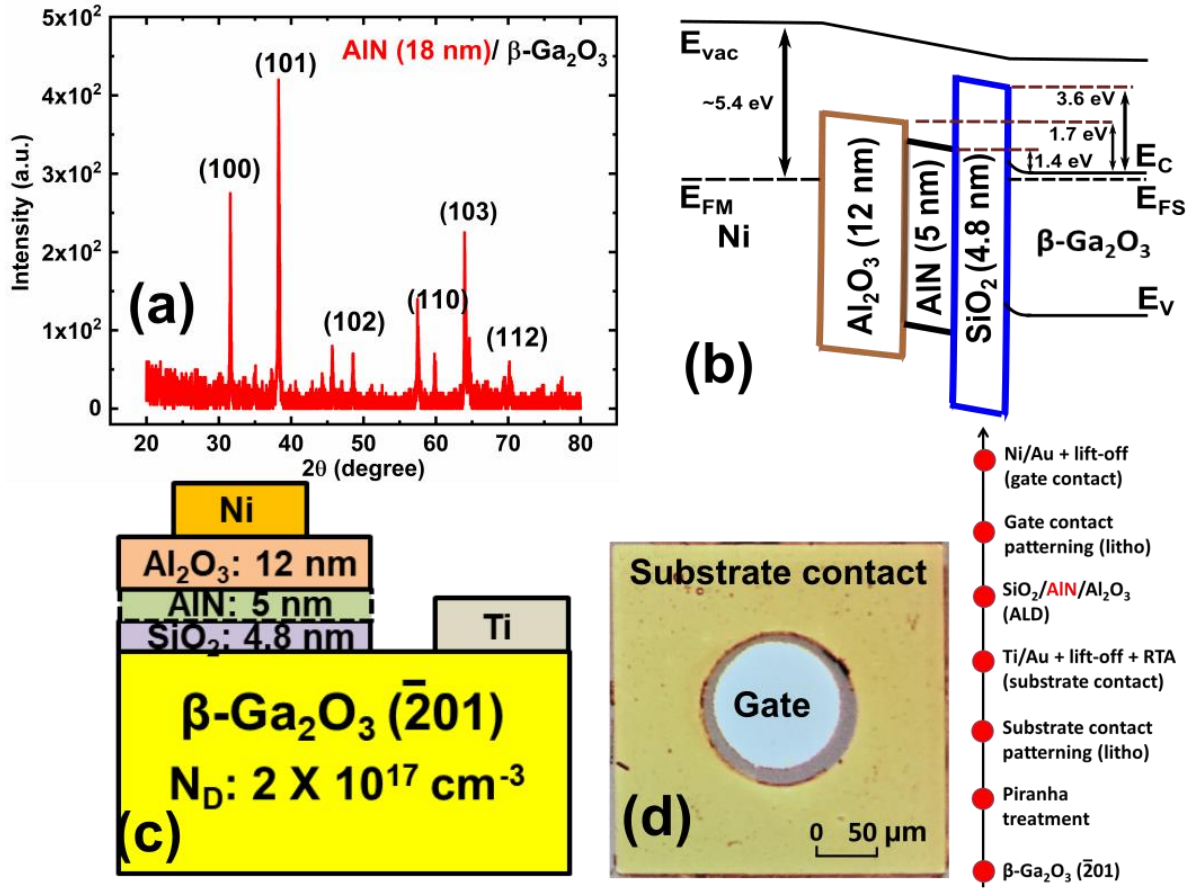

FIG. 1. (a) Grazing incidence x-ray diffraction pattern confirming the formation of polycrystalline AlN on $\beta$-Ga$_2$O$_3$. (b) Electron energy band diagram of Ni/Al$_2$O$_3$/AlN/SiO$_2$/$\beta$-Ga$_2$O$_3$ sample at equilibrium. (c) Cross-section schematic and (d) optical micrograph of fabricated MOSCAPs with a step-wise process flow.

For a comparative study, two 5×5 mm$^2$ square samples from the same Sn doped $\beta$-Ga$_2$O$_3$ ($\bar{2}$01) wafer with a background doping of 2×10$^{17}$ cm$^{-3}$ (Tamura corporation) were used for the fabrication of Ni/Al$_2$O$_3$/SiO$_2$/$\beta$-Ga$_2$O$_3$ and Ni/Al$_2$O$_3$/AlN/SiO$_2$/$\beta$-Ga$_2$O$_3$ MOSCAPs. The background doping concentration of the substrate was verified by fitting the $1/C^2 - V$ plot (shown in supplementary material) generated from the capacitance-voltage ($C - V$) data. At the start of the fabrication process, the samples were degreased organically by ultrasonication for 3 min each in acetone and methanol followed by a piranha solution (deionised (DI) water:30% H$_2$O$_2$:96% H$_2$SO$_4$ in 1:1:4 ratio) dip for 5 min. This was done to ensure a high quality SiO$_2$/$\beta$-Ga$_2$O$_3$ interface, as previously reported.[21] Optical lithography was used for defining the substrate contact region prior to ohmic metal (Ti/Au) deposition in an electron beam evaporator (EBE) and lift-off. The samples were rapid annealed at a temperature of 470 °C for 1 min in an $N_2$ atmosphere to lower the contact resistance. This was followed by simultaneous deposition of 4.8 nm thick $SiO_2$ on both samples by PEALD using tris(dimethylamino)silane and oxygen plasma at 250 °C. The control sample $Al_2O_3$/$SiO_2$ was unloaded from the ALD chamber and a 5 nm AlN charge trap layer was deposited on the other sample at 200 °C using TMA and ammonia plasma, before depositing 12 nm $Al_2O_3$ simultaneously on both samples using TMA and DI water at 250 °C in the same chamber. Thickness and stoichiometry of the gate dielectrics were studied using spectroscopic ellipsometry and XPS, respectively. To confirm the formation of $Al_2O_3$/AlN/$SiO_2$ and $Al_2O_3$/$SiO_2$ stacks, XPS depth profiles were measured on silicon monitor wafers loaded simultaneously in the ALD chamber during gate dielectric deposition (supplementary material). Next, optical lithography, metal deposition by EBE and lift-off were done to form Ni/Au gate contacts. Finally, the dielectrics were selectively etched from the top of the substrate contacts using a buffered-oxide-etch solution (5:1) after defining etch areas through optical lithography. Figs. 1(c) and (d) show the cross-section schematic and top view (with a step-wise process flow), respectively, of the fabricated MOSCAPs.

Fig. 2(a) shows the $C - V$ characteristics of as-fabricated Ni/$Al_2O_3$/$SiO_2$/$β$-$Ga_2O_3$ and Ni/$Al_2O_3$/AlN/$SiO_2$/$β$-$Ga_2O_3$ gate stacks measured using a B1500A semiconductor device analyzer. A peak-to-peak ac signal of 30 mV superimposed on a dc voltage swept at 50 mV/s from depletion to accumulation with an initial hold time of 10 s was used for frequency dependent $C - V$ measurements. Negligible frequency dispersion (supplementary material) combined with steep $C - V$ characteristics indicate a high quality $SiO_2$/$β$-$Ga_2O_3$ interface. Along with an increase in effective oxide thickness (EOT) as seen in the reduced accumulation capacitance ($C_{ox}$), the incorporation of AlN between $Al_2O_3$ and $SiO_2$ reduces the effect of positive oxide charge in $Al_2O_3$ on $β$-$Ga_2O_3$, thus increasing the $V_{fb}$ w.r.t Ni/$Al_2O_3$/$SiO_2$/$β Ga_2O_3$. Further, application of a positive voltage stress (charging voltage pulse) above 6 V increases the $V_{fb}$ in both gate stacks (Figs. 2(b) and (c)), attributed to the trapping of electrons injected through the thin $SiO_2$ interfacial layer into the gate dielectric (AlN or $Al_2O_3$) by F-N tunneling, as shown in Fig. 2(d). Details about the charge injection process are provided in the supplementary material. For the Ni/$Al_2O_3$/$SiO_2$/$\beta$-$Ga_2O_3$ stack, probable sites for electron trapping are located at the $Al_2O_3$/$SiO_2$ interface. On the other hand, charge trapping in deep level trap states at the polycrystalline grain boundaries in AlN are likely to be responsible for $V_{fb}$ shifts till 10.6 V in the Ni/$Al_2O_3$/AlN/$SiO_2$/$\beta$-$Ga_2O_3$ stack.[25,26]

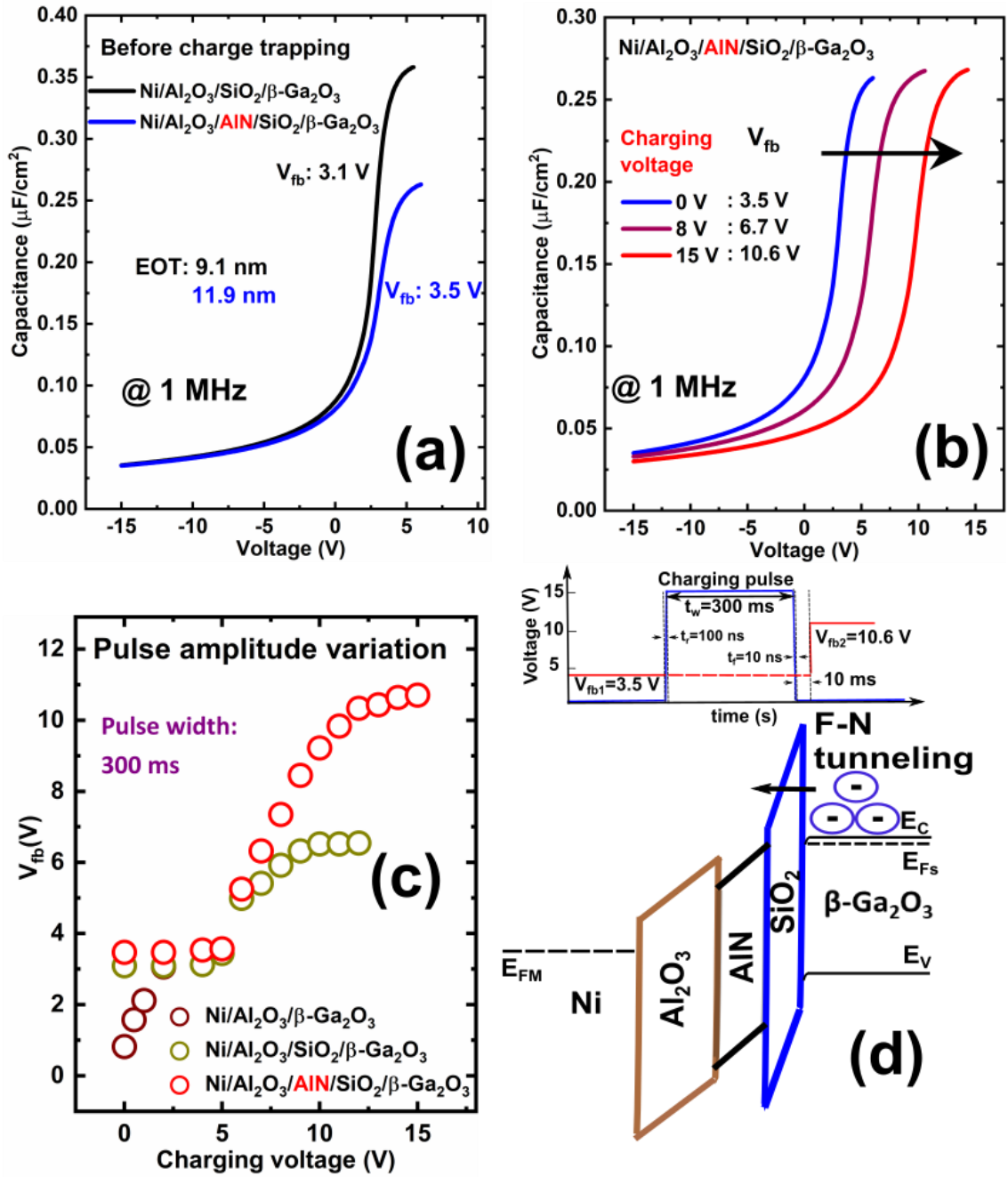

FIG. 2. $C-V$ characteristics of (a) as-fabricated Ni/Al$_2$O$_3$/SiO$_2$/$\beta$-Ga$_2$O$_3$ and Ni/Al$_2$O$_3$/AlN/SiO$_2$/$\beta$-Ga$_2$O$_3$ gate stacks before charge trapping, and, (b) Ni/Al$_2$O$_3$/AlN/SiO$_2$/$\beta$-Ga$_2$O$_3$ gate stack after applying 8 and 15 V charging voltage pulses. (c) $V_{fb}$ tunability with varying amplitude of charging voltage pulse. (d) Electron energy band diagram showing tunneling of electrons into the gate stack under positive voltage stress applied by a pulsed signal (timing diagram shown above).

The presence of deep level trap states in polycrystalline AlN is supported by the excellent retention of $V_{fb}$ observed in Ni/$Al_2O_3$/AlN/$SiO_2$/$\beta$-$Ga_2O_3$ as shown in Figs. 3(a) and (b). After $10^4$ s at 55 °C, a $V_{fb}$ reduction of only 0.3 V (0.1 V at room temperature) was observed in contrast to 2.6 V (0.7 V at room temperature) for Ni/$Al_2O_3$/$SiO_2$/$\beta$-$Ga_2O_3$. Hysteresis measurements using bi-directional $C - V$ sweeps were carried out before and after charge trapping (supplementary material) which ruled out the impact of border trap occupancy on $V_{fb}$ tunability. A gradual increase in $V_{fb}$ was observed in Ni/$Al_2O_3$/$\beta$-$Ga_2O_3$ MOSCAPs also (Fig. 2(c)). This can be attributed to the occupancy of linearly distributed (in space and energy) trap states in $Al_2O_3$.[27] However, very poor $V_{fb}$ retention of 2.2 V at room temperature was observed for the Ni/$Al_2O_3$/$\beta$-$Ga_2O_3$ stack.

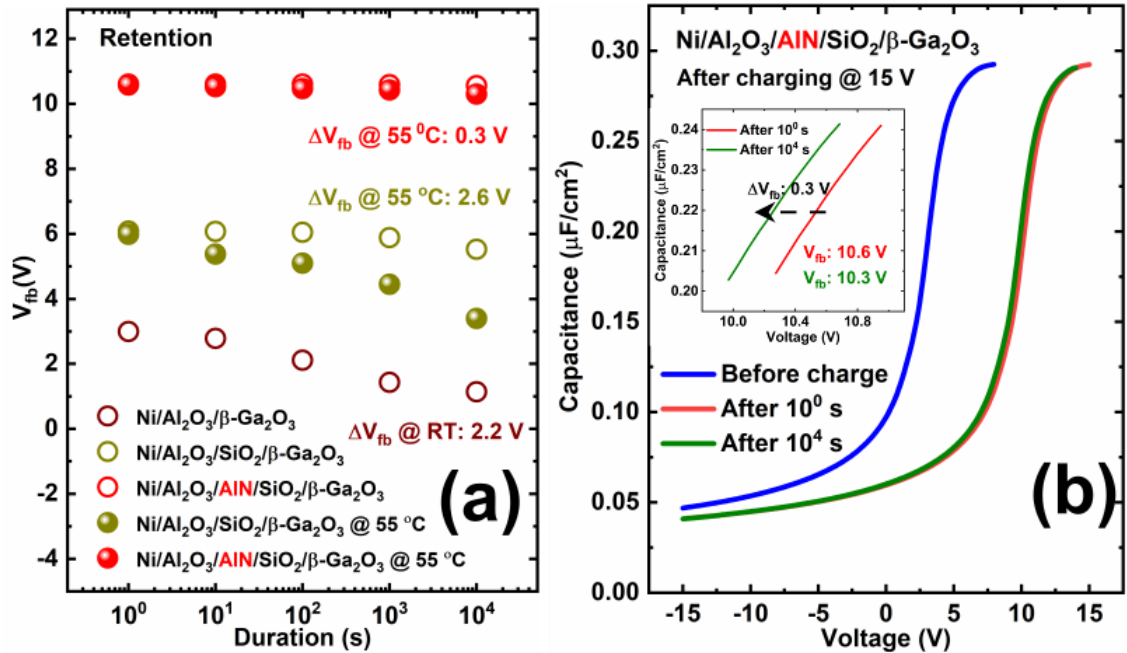

FIG. 3. (a) $V_{fb}$ retention w.r.t time at room temperature and 55 °C for various gate dielectric stacks, and, (b) $C-V$ characteristics of Ni/$Al_2O_3$/AlN/$SiO_2$/$\beta$-$Ga_2O_3$ MOSCAP showing the effect of charge trapping on $V_{fb}$ followed by its negligible (~0.3 V) reduction after $10^4$ s at 55 °C.

As shown in Figs. 4(a) and (b), nearly the same amount of charge can be modulated before and after charge trapping by both gate stacks throughout their $V_{fb}$ range. Charge values were extracted by integrating the area under their respective $C$–$V$ curves.

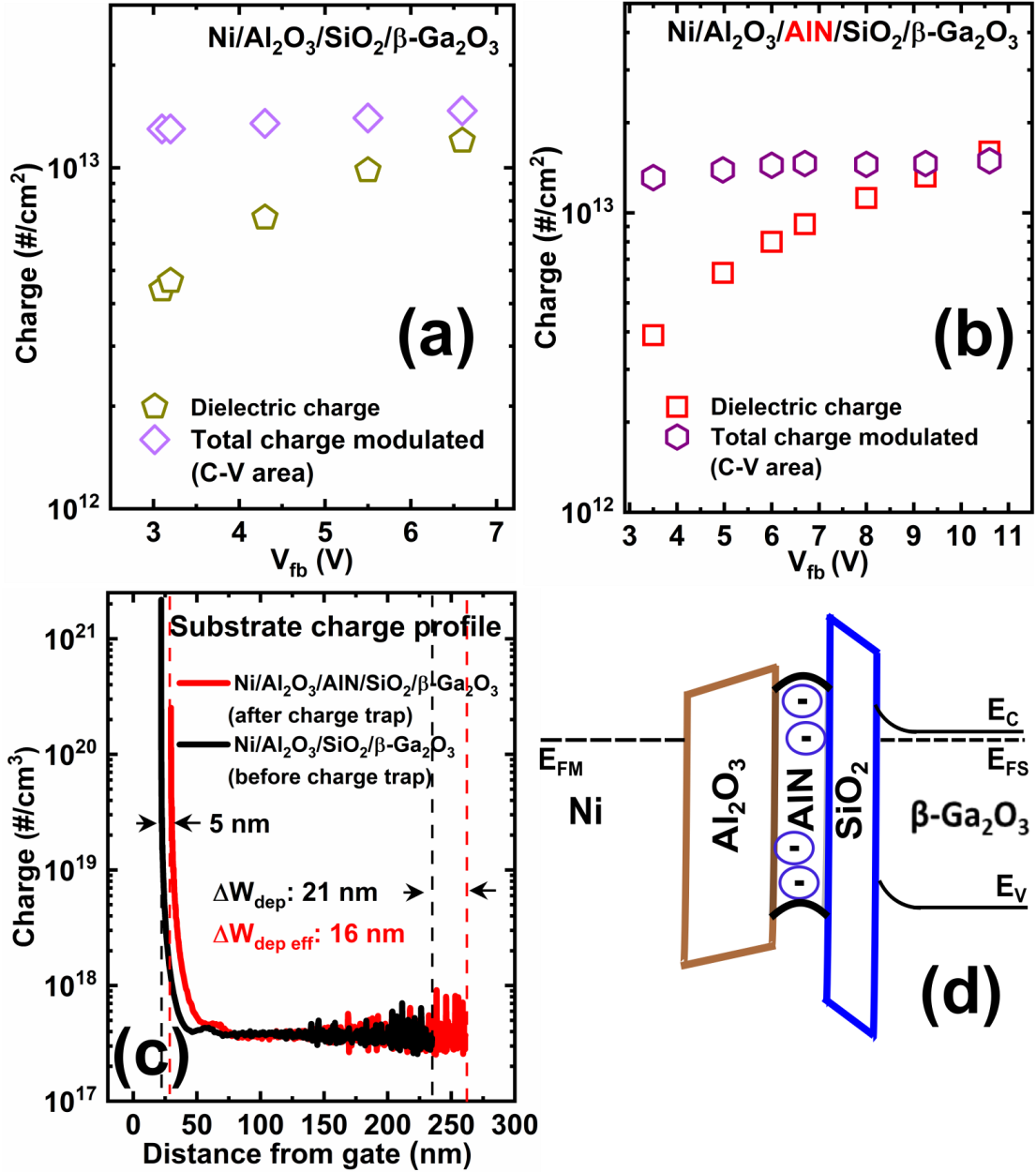

FIG. 4. Plots showing stored dielectric charge and $C$–$V$ modulated substrate charge densities for (a) Ni/Al$_2$O$_3$/SiO$_2$/$\beta$-Ga$_2$O$_3$, and, (b) Ni/Al$_2$O$_3$/AlN/SiO$_2$/$\beta$-Ga$_2$O$_3$ gate stacks. (c) Substrate charge profiles of Ni/Al$_2$O$_3$/SiO$_2$/$\beta$-Ga$_2$O$_3$ (before charge trapping) and Ni/Al$_2$O$_3$/AlN/SiO$_2$/$\beta$Ga$_2$O$_3$ (after charge trapping), and, (d) equilibrium electron energy band diagram of Ni/Al$_2$O$_3$/AlN/SiO$_2$/$\beta$-Ga$_2$O$_3$ (after charge trapping).

Despite a lower $C_{ox}$ (due to higher EOT) the Ni/Al$_2$O$_3$/AlN/SiO$_2$/β-Ga$_2$O$_3$ stack is able to modulate nearly the same amount of charge as the Ni/Al$_2$O$_3$/SiO$_2$/β-Ga$_2$O$_3$ stack due to extra substrate charge depletion as observed through the increase in depletion width (Fig. 4(c)). The increase in $V_{fb}$ is due to added negative dielectric charge injected under the application of a positive voltage pulse. The effective dielectric charge density ($N_{eff}$) was calculated using equation 1 and the $V_{fb}$ value extracted from $C-V$ characteristics,

$$N_{eff} = \frac{C_{ox}|\phi_{MS} - V_{fb}|}{qA} \quad (1)$$

where $\phi_{MS}$ is the work function difference between the gate metal (Ni) and β-Ga$_2$O$_3$, $q$ is unit electric charge and $A$ is the area of the fabricated MOSCAPs.

Reduced gate leakage current density, $J_g$, for high negative gate voltages exhibited by the Ni/Al$_2$O$_3$/AlN/SiO$_2$/β-Ga$_2$O$_3$ capacitor is likely due to better electric field distribution in the dielectric stack in comparison to the Ni/Al$_2$O$_3$/SiO$_2$/β-Ga$_2$O$_3$ sample. Higher EOT due to the introduction of AlN in Ni/Al$_2$O$_3$/AlN/SiO$_2$/β-Ga$_2$O$_3$ gate stack further helps in reducing $J_g$ for higher negative as well as positive gate voltages in comparison to the Ni/Al$_2$O$_3$/SiO$_2$/β-Ga$_2$O$_3$ capacitor. After charge trapping, an increase in $J_g$ for higher negative gate voltages (< -25 V) was seen for the Ni/Al$_2$O$_3$/SiO$_2$/β-Ga$_2$O$_3$ sample (Fig. 5(a)). In contrast, the Ni/Al$_2$O$_3$/AlN/SiO$_2$/β-Ga$_2$O$_3$ sample demonstrated a consistently low $J_g$ till $V_g$ of ~-60 V before as well as after charge trapping (Fig. 5(b)). Hence, apart from fine tuning $V_{fb}$ for normally-off operation, the charge trap stack is also capable of demonstrating higher breakdown voltages compared to conventional bilayer gate stacks. Table 1 summarizes the positive $V_{fb}$ and $V_t$ values for gate stacks and enhancement mode transistors, respectively, obtained using different material, process and design techniques.

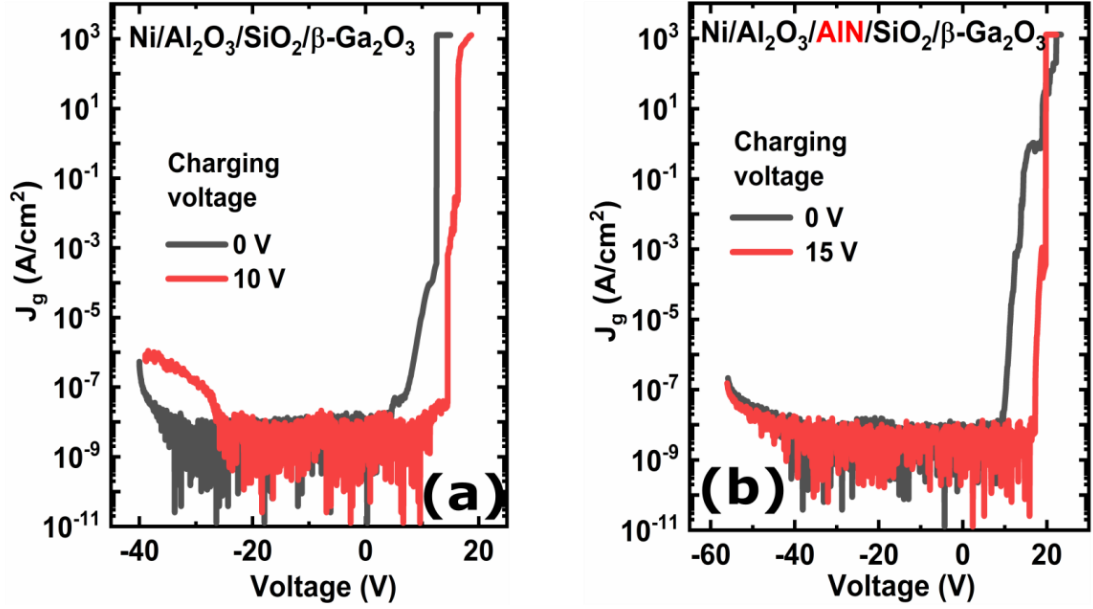

FIG. 5. (a) $J_g$ vs $V_g$ for (a) Ni/Al$_2$O$_3$/SiO$_2$/$\beta$-Ga$_2$O$_3$, and, (b) Ni/Al$_2$O$_3$/AlN/SiO$_2$/$\beta$-Ga$_2$O$_3$ gate stacks.

TABLE I. Summary of enhancement mode $V_t$ and $V_{fb}$ values for reported $\beta$-Ga$_2$O$_3$ MOSFETs and MOSCAPs, respectively.

| Ref no. | Method | $V_{fb}$(V) | $V_t$(V) |
| --- | --- | --- | --- |
| 17 | Gate recess | - | 0 |
| 18 | Wrap-gate fin FET | - | 1 |
| 19 | UID channel | - | 0 |
| 20 | Ferroelectric charge storage | - | 0 |
| 21 | LPCVD+PDA | 3.5 | - |
| 21 | ALD+FGA | 2.5 | - |
| 24 | Bi-layer dielectric | 3.25 | - |
| **This work** | **Charge trap layer** | **3.5 - 10.6** | - |

In summary, we demonstrate positive and tunable $V_{fb}$ with superior breakdown properties using a polycrystalline AlN charge trap inter-layer between Al$_2$O$_3$ and SiO$_2$ for enhancement mode $\beta$-Ga$_2$O$_3$ transistors. Positive $V_{fb}$ ranging from 3.5 V to 10.6 V (3× tunable) was realized without compromising the dielectric/channel interface quality. Excellent $V_{fb}$ retention of ∼97% measured till $10^4$ s at 55 °C after charge trapping demonstrates robust gate stack operation with less frequent requirement of charge injection cycles.

See supplementary material for details on extraction of *β*-$Ga_2O_3$ doping density, charge trapping mechanism, XPS analyses confirming the formation of AlON along with depth profiles for the $Al_2O_3$/AlN/$SiO_2$ and $Al_2O_3$/$SiO_2$ stacks.

The authors acknowledge the Ministry of Electronics and Information Technology and Department of Science and Technology, Government of India, for funding this work.

# Supplementary Material

# Charge trap layer enabled positive tunable $V_{fb}$ in β-Ga$_2$O$_3$ gate stacks for enhancement mode transistors

Dipankar Biswas,[1] Chandan Joishi,[1,2] Jayeeta Biswas,[1] Prabhans Tiwari,[1] and

Saurabh Lodha[1,a]

[1]Department of Electrical Engineering, Indian Institute of Technology Bombay, Mumbai, Maharashtra 400076, India

[2]Department of Electrical and Computer Engineering, The Ohio State University, Columbus, OH 43210, U.S.A.

E-mail: slodha@ee.iitb.ac.in

[a]To whom correspondence should be addressed

# Extraction of $\beta$-Ga$_2$O$_3$ substrate doping concentration using reverse-biased *C-V* characteristics

Background doping ($N_D$) of the $\beta$-Ga$_2$O$_3$ substrate used in this work has been extracted using the standard technique of reverse-biased capacitance-voltage *C-V* measurement. Fig. 1(a) shows the *C–V* characteristics of an as-fabricated Ni/Al$_2$O$_3$/SiO$_2$/$\beta$-Ga$_2$O$_3$ capacitor. From the slope of the $1/C^2$ –*V* plot (shown in Fig. 1(b)), obtained from the *C-V* data shown in Fig. 1(a), $N_D$ has been extracted using, $N_D = 2/[q\varepsilon_r\varepsilon_o A^2 d(C^2)/dV]$ where, $q$, $\varepsilon_r$, $\varepsilon_o$ and $A$ are elementary charge, relative permittivity of $\beta$-Ga$_2$O$_3$, permittivitty of vacuum and gate area, respectively.

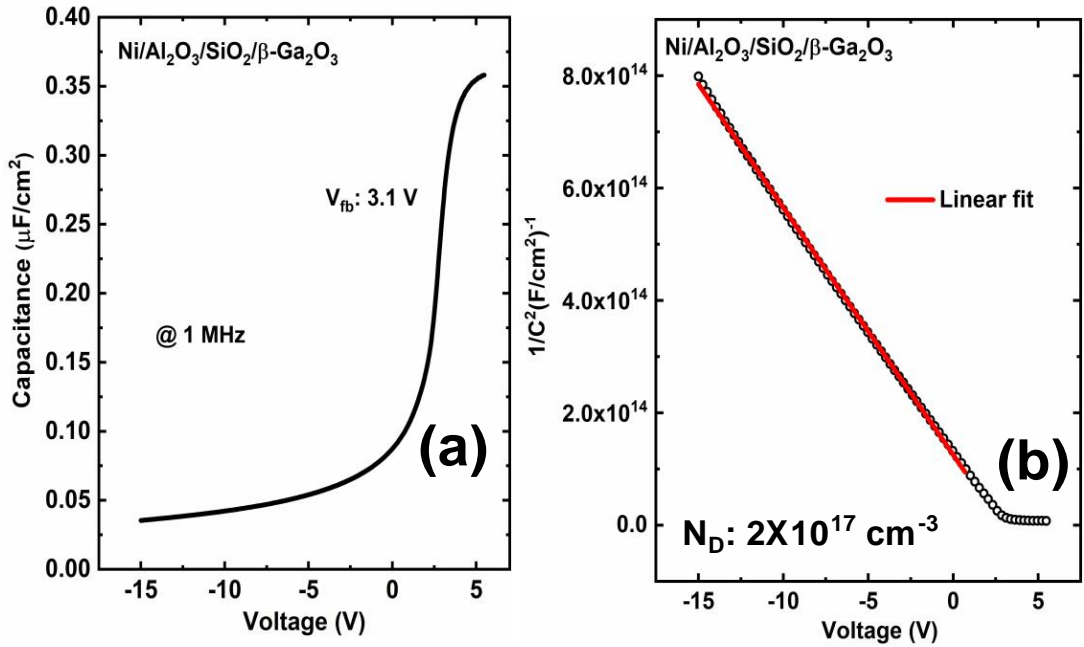

Figure 1: (a) *C–V* characteristics of as-fabricated Ni/Al$_2$O$_3$/SiO$_2$/$\beta$-Ga$_2$O$_3$ gate stack, (b) $1/C^2$–*V* plot used to extract $N_D$.

# Charge injection mechanism in the Ni/Al$_2$O$_3$/SiO$_2$/$\beta$-Ga$_2$O$_3$ and Ni/Al$_2$O$_3$/AlN/SiO$_2$/$\beta$-Ga$_2$O$_3$ gate stacks

Fig. 2(a) shows the variation in the gate current density $J_g$ with increasing positive stress voltage applied at the gate of a Ni/Al$_2$O$_3$/AlN/SiO$_2$/$\beta$-Ga$_2$O$_3$ MOSCAP. The linear dependence of $\ln J_g/E_{ox}^2$ on $1/E_{ox}$ (as shown by the red line in Fig. 2(b)) confirms F-N tunneling to be the dominant mechanism for conduction through the gate stack beyond 6 V of gate bias.[1,2] $E_{ox}$ is the electric field across 4.8 nm SiO$_2$.

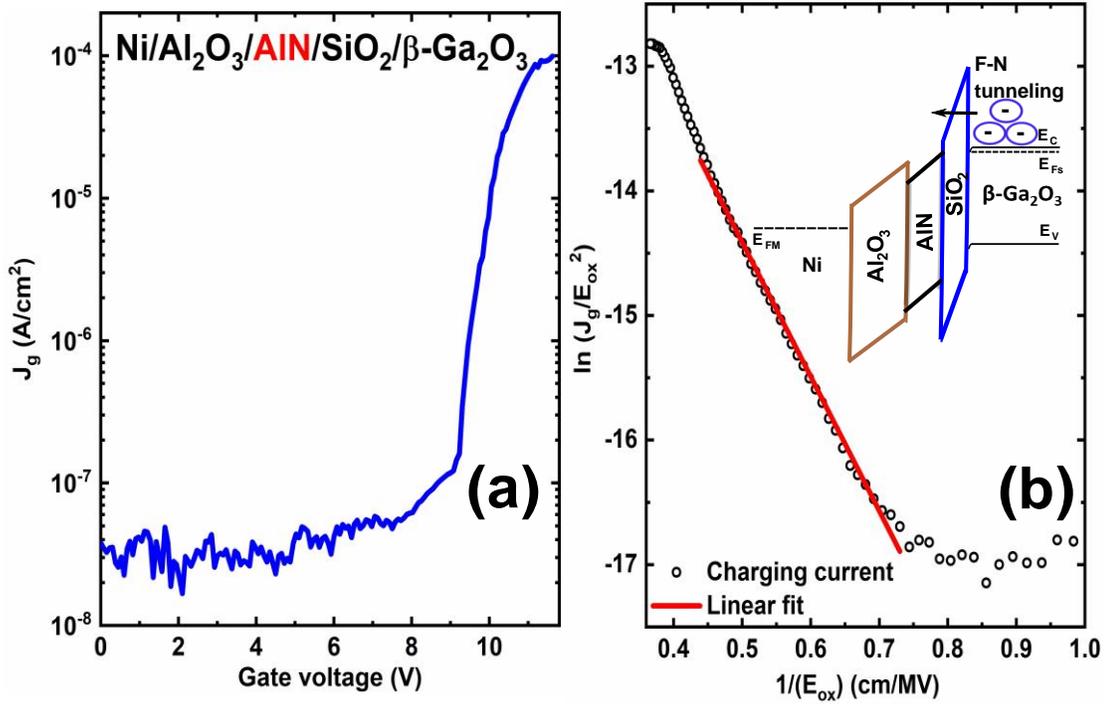

Figure 2: (a) $J_g$ vs gate voltage for the Ni/Al$_2$O$_3$/AlN/SiO$_2$/β-Ga$_2$O$_3$ capacitor, (b) ln $J_g/E_{ox}^2$ vs $1/E_{ox}$ obtained from $J_g$ vs gate voltage shown in Fig. 2(a). The red line shows the linear dependence of ln $J_g/E_{ox}^2$ on $1/E_{ox}$ confirming F-N tunneling to be the charge injection mechanism.

## Border trap analysis and its impact on $V_{fb}$ after charge trapping in the gate stack

Bi-directional HFCV characteristics measured at a frequency of 1 MHz (after charge trapping) highlight $V_{fb}$ shifts of 0.1 V and 0.3 V. These lead to border trap densities of 7.1×10$^{11}$ cm$^{-2}$ and 1.6×10$^{12}$ cm$^{-2}$ for Ni/Al$_2$O$_3$/SiO$_2$/β-Ga$_2$O$_3$ and Ni/Al$_2$O$_3$/AlN/SiO$_2$/β-Ga$_2$O$_3$ gate stacks, respectively. Hence, the large positive shifts in $V_{fb}$ along with its tunability cannot be attributed to border traps.

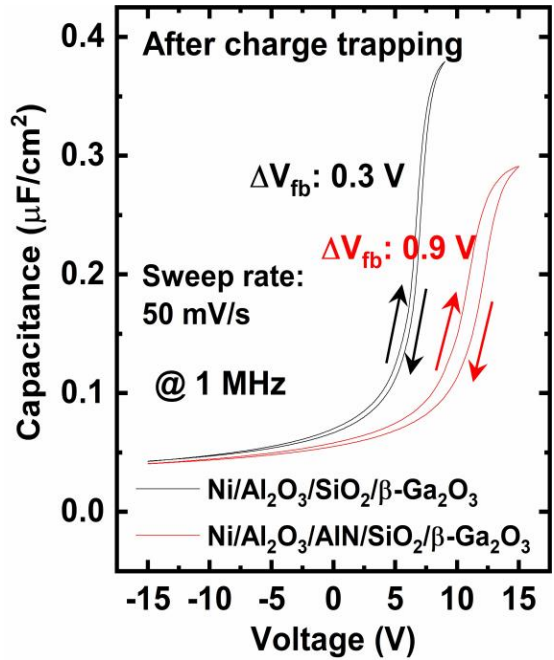

Figure 3: Bi-directional $C - V$ hysteresis data for Ni/Al$_2$O$_3$/SiO$_2$/$\beta$-Ga$_2$O$_3$ and Ni/Al$_2$O$_3$/AlN/SiO$_2$/$\beta$-Ga$_2$O$_3$ gate stacks measured after charge trapping.

## XPS analysis of AlN deposited by PEALD at 200 °C

Fig. 4(a) shows the de-convoluted Al2p XPS spectrum where the main peak position of 74.41 eV confirms the presence of Al-N bonds (i.e. AlN), whereas, the sub-peak at 75.61 eV corresponds to Al-O bonds, indicating the presence of Al$_x$O$_y$ (where x<2, y<3). This is consistent with the N1s spectrum shown in Fig. 4(b) where the presence of a main peak at 397.26 eV corresponds to Al-N bonds whereas, the main sub-peak at 398.73 eV suggests the formation of N-Al-O bonds (i.e. AlON). A lower energy sub-peak at 395.77 eV is due to N-N bonds. The atomic percentages extracted from the areas under the Al2p, N1s and O1s (Fig. 4(c)) spectra show a distribution of 39.3%, 35.9% and 24.7% for Al, N and O in the deposited AlN film.[3,4]

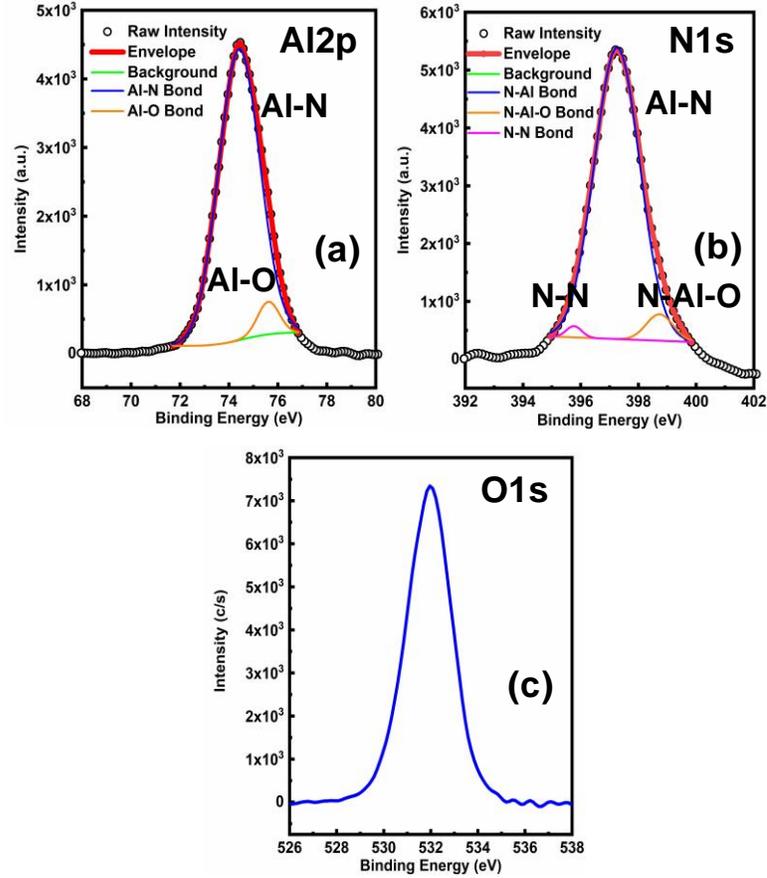

Figure 4: (a) Al2p, (b) N1s and (c) O1s XPS spectra of AlN dielectric.

To verify the formation of $Al_2O_3/SiO_2$ and $Al_2O_3/AlN/SiO_2$ stacks on $\beta$-$Ga_2O_3$, XPS depth profile measurements were performed on Si monitor wafers (loaded in the ALD chamber during gate dielectric stack deposition on $\beta$-$Ga_2O_3$). Figs. 5(a) and (b) show the depth profiles of $Al_2O_3$ (12 nm)/$SiO_2$ (4.8 nm) and $Al_2O_3$ (12 nm)/AlN (5 nm)/$SiO_2$ (4.8 nm) dielectric stacks on Si, respectively. The presence of AlN (AlON) between 12 nm $Al_2O_3$ and and 4.8 nm $SiO_2$ has been highlighted in Fig. 5(b).

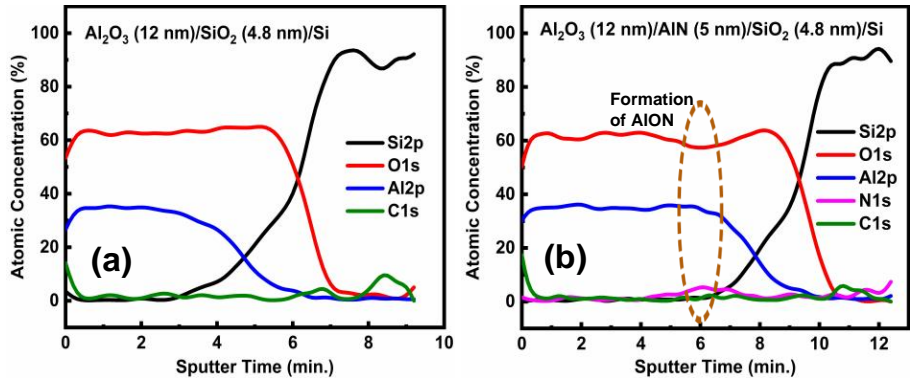

Figure 5: XPS depth profiles of (a) Al$_2$O$_3$/SiO$_2$/Si and (b) Al$_2$O$_3$/AlN/SiO$_2$/Si stacks.